\documentclass[aps,prx,reprint,superscriptaddress,longbibliography,floatfix]{revtex4-2}
\usepackage{amsmath,amssymb,graphicx,bm,xcolor,float}
\usepackage[colorlinks=true,urlcolor=blue,citecolor=blue,linkcolor=blue]{hyperref}
\usepackage{multirow}
\begin{document}

% \title{Thermodynamic-limit bootstrap: Solving fractional quantum Hall ground states}
\title{Solving fractional quantum Hall problems with thermodynamic-limit bootstrap}

\author{Qiang Gao}
\email{qianggao@mit.edu}
\affiliation{Department of Physics, Massachusetts Institute of Technology, Cambridge, MA 02139, USA}

\author{Taige Wang}
\email{tgwang@mit.edu}
\affiliation{Materials Research Laboratory, Massachusetts Institute of Technology, Cambridge, MA 02139, USA}
\affiliation{Department of Physics, Harvard University, Cambridge, MA 02138, USA}

\begin{abstract}
Determining correlated ground states directly in the thermodynamic limit remains a central challenge in quantum many-body physics. Here we develop a quantum many-body bootstrap that yields a rigorous and tight lower bound on the thermodynamic ground-state energy of a fractional quantum Hall system together with an optimized reduced density matrix $\rho_W^\star$ on a spatial window. For sufficiently tight bounds, we find that $\rho_W^\star$ accurately reproduces the ground-state reduced density matrix, as directly tested by its fidelity with known model states. For exactly solvable Laughlin and Moore--Read parent Hamiltonians, the bootstrap also reproduces pair correlations and orbital entanglement spectra (OES). For Landau-level-projected Coulomb interactions across multiple fillings, the energy bounds closely track established thermodynamic estimates, while the window density matrix recovers the familiar $\nu=1/3$ Laughlin and $\nu=5/2$ Moore--Read states and the composite Fermi liquid at $\nu=1/2$. More importantly, the same framework also resolves unsettled regimes. At $\nu=1/4$, the OES and real-space correlations identify a composite Fermi liquid, while at $\nu=12/5$ the combined diagnostics suggest $\overline{\mathrm{RR}}_3$ as the ground-state order. Our bootstrap therefore provides direct access to both ground-state energies and many-body structure in the thermodynamic limit.
\end{abstract}

\maketitle

\begin{figure}[h!]
\centering
\includegraphics[width=\columnwidth]{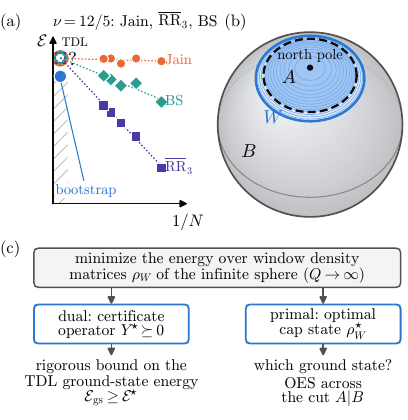}
\caption{\textbf{Thermodynamic-limit bootstrap as a ground-state solver.}
(a) Energy per particle $\mathcal{E}$ at $\nu = 12/5$ versus inverse particle number $1/N_e$. Finite-size corrections obscure the competition among the Jain, $\overline{\mathrm{RR}}_3$ and Bonderson--Slingerland (BS) branches, whereas the TDL bootstrap works directly at $1/N_e=0$ without selecting a candidate state in advance.
(b) A polar window $W$ selects a finite region of the infinite fluid, while an inner cap $A\subset W$ defines the orbital cut $A|B$.
(c) Energy minimization yields both a rigorous lower bound $\mathcal{E}\geq\mathcal{E}^\star$ and a window density matrix $\rho_W^\star$, whose fidelity and orbital entanglement spectrum (OES) probe the ground-state structure.}
\label{fig:schematic}
\end{figure}

Solving a strongly interacting quantum system requires determining its ground state in the thermodynamic limit (TDL). Most numerical methods approach this limit through finite systems, whose competing states can change relative order with system size. Fractional quantum Hall (FQH) systems provide a natural setting for this problem. At $\nu=12/5$, for example, finite-size corrections obscure the energetic competition among the Abelian Jain state, the non-Abelian Bonderson--Slingerland (BS) state and the particle--hole-conjugate $\overline{\mathrm{RR}}_3$ state~\cite{BondersonSlingerland2008,RezayiRead2009,Zhu2015twelvefifths,Mong2017}, as illustrated in Fig.~\ref{fig:schematic}(a). Here we show that the quantum many-body bootstrap solves FQH ground states directly in the TDL.

Fractional quantum Hall systems are among the most extensively studied strongly correlated quantum systems, yet many ground states remain unsettled. Exact diagonalization (ED) has been central to identifying FQH phases~\cite{YoshiokaHalperinLee1983,Haldane1983,Morf1998,RezayiHaldane2000}; density-matrix renormalization group (DMRG) extends the accessible sizes~\cite{ShibataYoshioka2001,Feiguin2008}, and iDMRG reaches infinite cylinder length at finite circumference~\cite{ZaletelMongPollmann2013,ZaletelMongPollmannRezayi2015}. Trial states reach much larger systems~\cite{Laughlin1983,Jain1989,MooreRead1991,ReadRezayi1999}, but require the phase to be chosen in advance. Recently developed neural variational methods are particularly well suited to incorporating Landau-level mixing, but remain limited to system sizes comparable to ED~\cite{TengDaiFu2025,Qian2025DeepHall,AbouelkomsanFu2026MagNet,Gattu2026CFFlow}.

In this work, we address this problem with the quantum many-body bootstrap, which replaces a variational wavefunction by universal constraints on local correlations. Positivity, symmetry and consistency conditions on reduced density matrices turn the ground-state problem into a convex optimization via semidefinite programming (SDP)~\cite{Mazziotti2002,Mazziotti2004,BarthelHubener2012,Han2020bootstrap,NancarrowXin2023,FawziFawziScalet2024certified,ChadhaScheerKhalaf2026,gao2026bootstrapping,scheer2025defect}. For quantum Hall systems, one of us previously formulated a bootstrap based on the static structure factor, obtaining rigorous energy bounds and correlation signatures~\cite{Gao2025bootstrap}. That formulation still suffered from finite-size effects and unfavorable scaling, and its particle entanglement spectrum (PES) provided limited state discrimination. 

Accessing the TDL directly through SDP has long been a goal of the many-body bootstrap. Barthel and H{\"u}bener introduced an SDP framework for ground-state energy bounds in translation-invariant local systems directly in the TDL~\cite{BarthelHubener2012}. Rai et al. extended this to spectral-gap bounds for frustration-free spin systems~\cite{Rai2026hierarchyofspectral}. Recent progress toward the Laughlin gap conjecture~\cite{rougerie2019laughlin} includes two parallel approaches: a finite-orbital SDP hierarchy by Scheer and Khalaf~\cite{ScheerKhalaf2026} and a Fock-space construction by OpenAI~\cite{OpenAI2026Fock}, subsequently formulated as bootstrap by one of us~\cite{Gao2026LaughlinSDP}. Here, we instead target both the ground-state energy and local structure of FQH fluids directly in the TDL.

We work with an infinite system through a finite window $W$ (Fig.~\ref{fig:schematic}(b)). By constraining correlations within this window, the bootstrap gives a rigorous lower bound $\mathcal E^\star$ on the thermodynamic ground-state energy per particle, $\mathcal{E}_{\rm gs}\equiv\lim_{N\to\infty}E_{\rm gs}(N)/N$, and determines the ground-state structure through the optimized reduced density matrix $\rho_W^\star$ supported on $W$ (Fig.~\ref{fig:schematic}(c)). The results below support a simple practical picture,
\begin{equation}
    \mathcal E^\star \lesssim \mathcal E_{\rm gs},
    \qquad
    \rho_W^\star \simeq \rho_{{\rm gs},W},
\label{eq:energy_state_closeness}
\end{equation}
where the second relation is tested directly through fidelity to benchmark ground states and independently through observables derived from the window density matrix. Restricting $\rho_W^\star$ to an inner cap $A\subset W$ gives an OES that probes topological order~\cite{LiHaldane2008,Thomale2010}, while its two-particle correlations determine the real-space pair distribution $g(r)$.

The method works across exactly solvable models, established Coulomb states and unsettled ground states. For the Laughlin and Moore--Read parent Hamiltonians~\cite{Laughlin1983,TrugmanKivelson1985,MooreRead1991}, the bootstrap reproduces the exact ground-state energies and gives cap fidelities $0.999$ and $0.991$ on a $K'=8$ cap, alongside pair correlations and OES that closely match the model states. For Coulomb interactions across the LLL and LL1, rigorous TDL energy bounds from $K=16$ windows lie within $0.2$--$1.6\%$ of thermodynamic estimates from ED and DMRG extrapolations~\cite{Morf1998,Feiguin2008,ZhaoShengHaldane2011,ZaletelMongPollmannRezayi2015}. The optimized density matrix $\rho_W^\star$ then supplies increasingly specific tests: fidelity to candidate states, OES for gapped topological order and $g(r)$ for composite Fermi liquids. At $\nu=12/5$, Jain, BS and $\overline{\mathrm{RR}}_3$ remain close competitors in the TDL. The combined local diagnostics suggest $\overline{\mathrm{RR}}_3$ as the most likely ground-state order~\cite{BondersonSlingerland2008,ReadRezayi1999,RezayiRead2009}. At LLL $\nu=1/4$, the OES remains inconclusive but $g(r)$ identifies a composite Fermi liquid. The same bootstrap treats gapped Abelian and non-Abelian topological phases and gapless composite Fermi liquids without prior assumptions about the phase, with their distinct physics revealed by the window density matrix.

\section*{A window onto the infinite fluid}

In this work, we restrict to a single partially filled Landau level, where the kinetic energy is quenched and the many-body ground state is determined entirely by interactions. For a rotationally invariant two-body interaction, the projected Hamiltonian is specified by its Haldane pseudopotentials~\cite{Haldane1983},
\begin{equation}
    H=\sum_{m\,\mathrm{odd}}V_m P_m,
\label{eq:window_hamiltonian}
\end{equation}
where $P_m$ projects a pair onto relative angular momentum $m$. For the Coulomb problems, the pseudopotentials are fixed by the Landau-level form factor. For the Laughlin and Moore--Read benchmarks, we use the corresponding two- and three-body parent Hamiltonians~\cite{Laughlin1983,TrugmanKivelson1985,MooreRead1991}.

We solve this interacting quantum Hall problem directly in the thermodynamic limit by keeping the physical fluid infinite and resolving its correlations within a finite orbital window $W$ containing the $K$ orbitals closest to the north pole (Fig.~\ref{fig:schematic}(b)). These orbitals become the planar symmetric-gauge orbitals in the TDL~\cite{Haldane1983}. Increasing $K$ extends the resolved correlation range of the same infinite system.

The optimization variable is the window Fock-space density matrix $\rho_W$. Its $k$-particle RDMs (kRDM) $\Gamma_W^{(k)}$ decompose into center-of-mass (COM) angular-momentum sectors $M$. A homogeneous infinite fluid has the same relative correlations in every such sector, which we impose as (symbolically with actual form given in Methods)
\begin{equation}
\Gamma_{W,M}^{(k)}
=
C_{W,M}^{(k)}
\gamma_{W}^{(k)}
C_{W,M}^{(k)\dagger},
\qquad
\gamma_{W}^{(k)}\succeq0 ,
\label{eq:window_homogeneity}
\end{equation}
for $k=2,3,4$ and all allowed $M$. Here the RDM $\gamma_{W}^{(k)}$ depends only on the relative angular momenta, and $C_{W,M}^{(k)}$ embeds it into the orbital basis with COM angular momentum $M$.

For two particles, $\gamma^{(2)}$ is related to the pair weights $\pi_m\geq0$ in the relative-angular-momentum channels $m$, as detailed in Methods~\cite{Gao2025bootstrap}. A window of $K$ orbitals constrains the channels $m\leq2K-3$. The complete thermodynamic two-body correlations are described by the static structure factor $S_q=\lim_{N\to\infty}N^{-1}\langle\delta\rho_{-\mathbf q}\delta\rho_{\mathbf q}\rangle$, with $S_q\geq0$, and the interaction energy is a linear functional of $S_q$. The resolved $\pi_m$ fix the corresponding Laguerre moments of $S_q$~\cite{GMP1986,Gao2025bootstrap}, while the unresolved moments remain variational. Thus the finite window constrains the thermodynamic two-body correlations without truncating them. Minimizing over the unresolved correlations gives a rigorous lower bound to the full TDL energy.

These ingredients define the thermodynamic bootstrap,
\begin{equation}
\begin{aligned}
\mathcal{E}^\star
&=\min_{\rho_W,S_q}\mathcal{E}[S_q]
\\
\text{s.t.}\quad
&\rho_W\succeq0\;\text{(positivity)},\qquad
{\rm Tr}\rho_W=1\;\text{(norm.)},
\\
&{\rm Tr}(\rho_W c_a^\dagger c_a)=\nu
\;\text{(filling)},
\\
&\Gamma_{W,M}^{(k)}
=
C_{W,M}^{(k)}
\gamma_{W}^{(k)}
C_{W,M}^{(k)\dagger},
\qquad
\gamma_{W}^{(k)}\succeq0
\;\text{(homo.)},\\
&S_q\geq0\;\text{(density fluctuation positivity)}.
\end{aligned}
\label{eq:window_bootstrap}
\end{equation}
For density-density interactions, the energy depends on the thermodynamic 2RDM through $S_q$, while $\rho_W$ constrains the allowed $S_q$. For the three-body Moore--Read parent, the energy depends directly on the window 3RDM.

The same optimization gives both the energy bound and the ground-state structure within the window through $\rho_W^\star$. The RDMs of $\rho_W^\star$ determine local correlations, and tracing $\rho_W^\star$ to an inner cap $A\subset W$ gives the OES across $A|B$ (Fig.~\ref{fig:schematic}(b))~\cite{LiHaldane2008,Thomale2010}. We show below that $\rho_W^\star$ reproduces the physical ground-state structure across independent diagnostics, even though global representability is not imposed.

The energy lower bound can be made rigorous through the dual SDP certificate~\cite{OpenAI2026Fock}. It produces a positive slack operator $Y_W^\star$ on the window, whose $SU(2)$ rotation average is a global positive operator~\cite{OpenAI2026Fock},
\begin{equation}
\int dg\,U_gY_W^\star U_g^\dagger\succeq0 .
\label{eq:window_certificate}
\end{equation}
Together with other dual positivity conditions, this certifies
$\mathcal{E}\geq\mathcal{E}^\star$ as detailed in Methods. Every finite-$K$ result therefore holds directly in the thermodynamic limit, without extrapolation in particle number or window size.

\begin{figure}[!tbp]
\centering
\includegraphics[width=0.92\columnwidth]{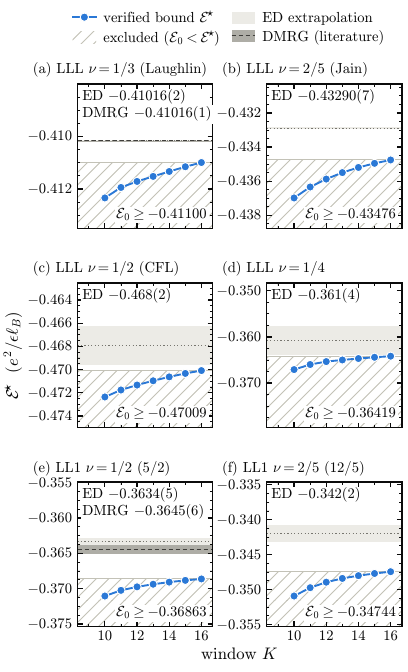}
\caption{\textbf{Certified thermodynamic Coulomb-energy bounds.}
Bootstrap lower bounds $\mathcal E^\star$ versus window size $K$ for
(a) LLL $\nu=1/3$,
(b) LLL $\nu=2/5$,
(c) LLL $\nu=1/2$,
(d) LLL $\nu=1/4$,
(e) LL1 $\nu=1/2$ ($5/2$) and
(f) LL1 $\nu=2/5$ ($12/5$).
LLL and LL1 denote the lowest and first excited Landau levels, respectively. Every finite-$K$ point is a thermodynamic-limit bound certified using rational arithmetic. The hatched region is excluded by $\mathcal E_{\rm gs}\geq\mathcal E^\star$, and each panel quotes the strongest bound at the largest $K$ shown. Shaded bands are our finite-sphere ED thermodynamic estimates, obtained from density-rescaled energies for systems with $N_e=10$--$18$ followed by linear extrapolation in $1/N$. Their widths show the sensitivity to leave-one-out fits and alternative parity or shift branches rather than statistical uncertainty. Published DMRG estimates are shown where available~\cite{ZhaoShengHaldane2011,ZaletelMongPollmannRezayi2015}.}
\label{fig:energy}
\end{figure}

\section*{TDL energy without finite-size extrapolation}

Energy provides the first quantitative test because FQH systems have vast number of thermodynamic benchmarks. As expected for frustration-free parent Hamiltonians, the Laughlin $V_1$ and three-body Moore--Read optimizations recover the exact zero energy to numerical precision~\cite{Laughlin1983,TrugmanKivelson1985,MooreRead1991}. Now we turn to Coulomb interactions whose ground state energy is nontrivial.

The bootstrap lower bounds lie very close to thermodynamic estimates from finite-size ED and DMRG extrapolations. Across the six fillings in Fig.~\ref{fig:energy}, the $K=16$ bounds differ from the corresponding thermodynamic estimates by only $0.8\times10^{-3}$--$5.4\times10^{-3}\,e^2/(\epsilon\ell_B)$ (for reference, $0.2$--$1.6\%$ of the corresponding ground-state energy magnitude). Two extensively studied cases illustrate this accuracy. At LLL $\nu=1/3$, the bootstrap gives $\mathcal E^\star=-0.41100$, only $8.4\times10^{-4}\,e^2/(\epsilon\ell_B)$ below the iDMRG extrapolation $-0.410164(4)$~\cite{ZaletelMongPollmannRezayi2015}. At half filling of LL1, $\mathcal E^\star=-0.36863$ lies $4.1\times10^{-3}\,e^2/(\epsilon\ell_B)$ below the sphere-DMRG extrapolation $-0.3645(6)$~\cite{ZhaoShengHaldane2011}.

For comparison, finite-size sphere ED and DMRG reach $41$ and $66$--$70$ active Landau-level orbitals, respectively~\cite{ZhaoShengHaldane2011}, yet the largest-size DMRG energies remain $9\times10^{-3}$--$1.6\times10^{-2}\,e^2/(\epsilon\ell_B)$ from the thermodynamic estimates at LLL $\nu=1/3$ and LL1 $\nu=1/2$. Composite-fermion variational Monte Carlo reaches much larger systems and comes within $1\times10^{-4}$--$3\times10^{-4}\,e^2/(\epsilon\ell_B)$ of the thermodynamic estimates at LLL $\nu=1/3$ and $\nu=2/5$~\cite{JainKamilla1997,Pu2017}, but its accuracy is tied to a prescribed trial-state manifold that must be chosen for each candidate phase. The bootstrap instead uses the same optimization framework across all fillings studied, without selecting a candidate state in advance, and every finite-$K$ result is already a rigorous lower bound directly in the TDL.

Finite-size effects are particularly severe at LL1 $\nu=2/5$. The largest reported $\overline{\mathrm{RR}}_3$ sphere-DMRG calculation with $N_e=22$ electrons remains about $8.8\times10^{-3}\,e^2/(\epsilon\ell_B)$ from its extrapolated thermodynamic energy, whereas the $K=16$ bootstrap bound lies only about $5.4\times10^{-3}\,e^2/(\epsilon\ell_B)$ from the same reference~\cite{Zhu2015twelvefifths}. In this case, the bootstrap is closer to the extrapolated thermodynamic energy than the largest published sphere-DMRG result.

The residual dependence on $K$ measures finite correlation range rather than finite physical size. Extrapolating this dependence brings the estimates even closer to thermodynamic estimates from other methods (Extended Data Fig.~\ref{fig:energy_extrap}). This extrapolation is used only to study convergence, and every point in Fig.~\ref{fig:energy} is already a certified thermodynamic-limit bound.

\section*{Fidelity test of the ground state}

Overlaps between trial wavefunctions and exact-diagonalization ground states have long been used to assess candidate FQH states~\cite{RezayiRead2009,Bonderson2012}. Here we extend this comparison to a local density matrix in the thermodynamic limit. The energy bound tests the ground-state energy, while the optimized density matrix allows a direct test of the relation $\rho_W^\star\simeq\rho_{{\rm gs},W}$. For an inner cap $A\subset W$ containing $K'$ orbitals, we form $\rho_A^\star=\operatorname{Tr}_{W\setminus A}\rho_W^\star$ and compare it with the corresponding model-state density matrix $\rho_A^{\rm model}$ through the fidelity $F(\rho,\sigma)=[\operatorname{Tr}\sqrt{\sqrt{\rho}\sigma\sqrt{\rho}}]^2$, which reduces to the squared wavefunction overlap for pure states~\cite{Jozsa1994}. A fidelity close to unity therefore tests whether the bootstrap reproduces the ground-state density matrix within the resolved region.

\begin{table}[t]
\centering
\caption{\textbf{Model-state fidelities of the optimized window density matrix.}
Fidelities are evaluated on $K'=8$ caps. Superscript $5$ denotes $K'=5$.
The APf fidelity is identical to that of Pf for particle--hole-symmetric fillings.
Monte Carlo references (Jain, BS and CFL) have an approximate fidelity resolution of $0.02$, except for starred entries ($0.1$, see Methods). The window size for bootstrap is $K=16$.}
\label{tab:fidelity}
\renewcommand{\arraystretch}{1.13}
\setlength{\tabcolsep}{3.0pt}
\small
\begin{tabular}{@{}p{0.3\columnwidth} p{\dimexpr0.7\columnwidth-2\tabcolsep\relax}@{}}
\hline\hline
\textbf{Hamiltonian} & \textbf{Model-state fidelity $F$} \\
\hline
$V_1$ parent & Laughlin (0.999) \\
3-body parent & Pf (0.991) \\
LLL $\nu=1/3$ & Laughlin (0.978) \\
LLL $\nu=2/5$ & Jain (0.973)$^{5}$, $\overline{\mathrm{RR}}_3$ (0.916)$^{5}$, BS (0.855)$^{5}$ \\
LLL $\nu=1/2$ & Pf (0.753), Pf (0.894)$^{5}$, CFL (0.80)$^{5,*}$ \\
LLL $\nu=1/4$ & Pf (0.896), Pf (0.968)$^{5}$, CFL (0.991)$^{5}$ \\
LL1 $\nu=1/2$ & Pf (0.771), Pf (0.904)$^{5}$, CFL (0.63)$^{5,*}$ \\
LL1 $\nu=2/5$ & BS (0.961)$^{5}$, $\overline{\mathrm{RR}}_3$ (0.944)$^{5}$, Jain (0.816)$^{5}$ \\
\hline\hline
\end{tabular}
\end{table}

We compare the bootstrap density matrices with Laughlin~\cite{Laughlin1983}, Pfaffian~\cite{MooreRead1991,FaugnoBalramBarkeshliJain2019}, Jain~\cite{Jain1989,JainKamilla1997}, $\overline{\mathrm{RR}}_3$~\cite{ReadRezayi1999,RezayiRead2009}, Bonderson--Slingerland~\cite{BondersonSlingerland2008} and composite-Fermi-sea~\cite{RezayiRead1994} reference states (Table~\ref{tab:fidelity}). Their constructions, geometries and sizes are specified in Methods. The parent Hamiltonians provide the most direct benchmark because their ground states are known independently. Their near-unit fidelities persist beyond the $K'=8$ caps reported in the table: at $K'=12$, the fidelities remain $0.996$ for Laughlin and $0.948$ for Moore--Read, demonstrating agreement over a larger resolved region.

For Coulomb interactions, fidelity tests both the accuracy of the local density matrix and its preference among candidate states. At LLL $\nu=1/3$, the Laughlin fidelity of $0.978$ shows that the agreement extends beyond exactly solvable parent Hamiltonians. At LLL $\nu=2/5$, Jain is preferred over both non-Abelian candidates, with $F=0.973$ compared with $0.916$ for the closest competitor, $\overline{\mathrm{RR}}_3$, on the same $K'=5$ cap. At quarter filling, the composite-Fermi-sea fidelity reaches $0.991$, suggesting CFL behavior to be further examined below. Fidelity thus recovers established states and guides the diagnosis of less settled regimes.

At LL1 $\nu=2/5$, the $K'=5$ cap gives fidelities $0.961$, $0.944$ and $0.816$ with BS, $\overline{\mathrm{RR}}_3$ and Jain, respectively. These values make Jain less likely, but leave the two non-Abelian candidates unresolved because their fidelity difference lies within the approximately $0.02$ Monte Carlo resolution (Methods). We therefore turn to the OES, whose low-lying structure provides a more universal probe of topological order, to seek a distinction beyond microscopic wavefunction agreement.

\begin{figure*}[!tbp]
\centering
\includegraphics[width=\textwidth]{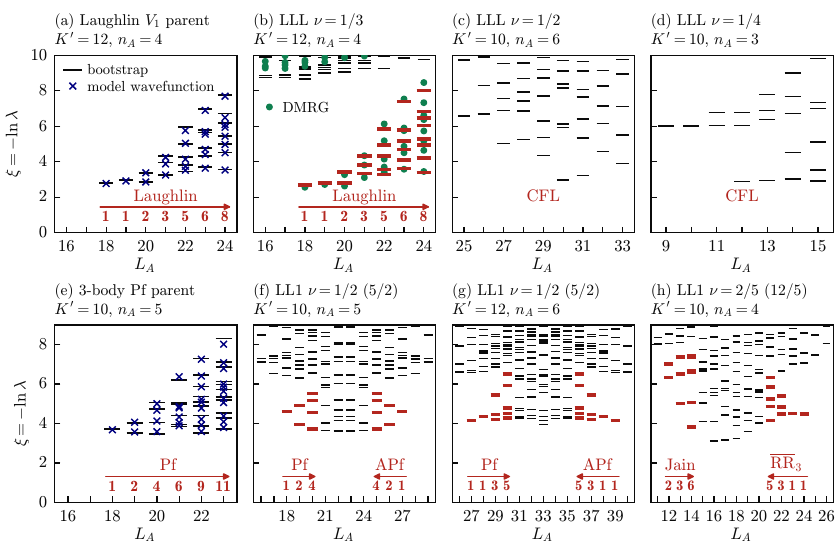}
\caption{\textbf{Orbital entanglement spectrum (OES) from bootstrap.}
Entanglement energies $\xi=-\ln\lambda$ for an inner cap of $K'$ orbitals, resolved by $n_A$ and $L_A$.
(a,b) The Laughlin parent and LLL $\nu=1/3$ exhibit the same Laughlin counting.
(c) LLL \(\nu=1/2\) shows no isolated low-lying chiral branch, as expected for a compressible composite Fermi liquid (CFL). (d) LLL \(\nu=1/4\) likewise shows no isolated Pfaffian-like branch. The OES disfavors simple Pfaffian order but does not by itself resolve the ground state.
(e) The three-body Moore--Read parent provides the Pfaffian (Pf) benchmark.
(f,g) LL1 $\nu=1/2$ ($5/2$) displays Pfaffian and anti-Pfaffian (APf) counting at opposite angular-momentum edges.
(h) LL1 $\nu=2/5$ ($12/5$) exhibits levels compatible with Jain counting $2,3,6$ and $\overline{\mathrm{RR}}_3$ counting $1,1,3,5$ at opposite angular-momentum edges. The finite-cap counting alone does not uniquely identify the order.
Red bars mark the counted low-lying levels. Blue crosses in (a,e) show the corresponding disk Laughlin and Moore--Read model states, while green circles in (b) show the same cut of an $N_e=24$ sphere-DMRG calculation at LLL $\nu=1/3$. The window size for bootstrap is $K=16$.}
\label{fig:gallery}
\end{figure*}

\section*{Topological order from the entanglement spectrum}

Fidelity compares the bootstrap density matrix with specific microscopic reference states. Topological phases also admit more universal diagnostics that probe topological data rather than microscopic wavefunction details. These include the orbital and particle entanglement spectra (OES and PES)~\cite{LiHaldane2008,Thomale2010,ChandranHermannsRegnaultBernevig2011,SterdyniakRegnaultBernevig2011}, topological entanglement entropy~\cite{KitaevPreskill2006,LevinWen2006} and momentum polarization~\cite{TuZhangQi2013,ZaletelMongPollmann2013}. Here we focus on the OES, whose low-lying structure provides a sensitive diagnostic of competing topological orders. Using the same inner cap $A$ of the first $K'$ orbitals, we define its complement $B=W\setminus A$ (Fig.~\ref{fig:schematic}(b)). The cap density matrix $\rho_A^\star=\operatorname{Tr}_B\rho_W^\star$ then gives the entanglement energies $\xi=-\ln\lambda$ through its eigenvalues $\lambda$, resolved by particle number $n_A$ and angular momentum $L_A$. The finite cap truncates the edge counting to the allowed $(K',n_A)$ configurations shown in Fig.~\ref{fig:gallery}.

For the parent Hamiltonians, the fidelity benchmark above directly establishes the closeness of the cap density matrices. The OES tests whether this agreement extends to their resolved entanglement structure. The Laughlin levels match nearly one by one, and the Moore--Read spectrum reproduces the same detailed branch structure with small level shifts. At LLL $\nu=1/3$, the Coulomb OES likewise recovers the Laughlin structure and agrees with the $N_e=24$ sphere-DMRG calculation, as shown in Fig.~\ref{fig:gallery}(b). Thus the same optimized density matrix that gives a tight energy bound and high fidelity also reproduces detailed entanglement structure of the ground state.

The PES characterizes the bulk quasihole sector and provides a complementary fingerprint of topological order~\cite{SterdyniakRegnaultBernevig2011,ChandranHermannsRegnaultBernevig2011}. Extended Data Fig.~\ref{fig:pes} reproduces the expected Laughlin and Moore--Read PES counting, but at LL1 $\nu=2/5$ the Jain and $\overline{\mathrm{RR}}_3$ candidates share the same low-lying four-hole PES counting. The OES contains more detailed edge information, although its finitized counting on a small cap need not uniquely identify a candidate state.

At half filling of LL1, the OES instead develops Moore--Read structure. Particle--hole symmetry maps
\begin{equation}
    (n_A,L_A)
    \longrightarrow
    \left(
    K'-n_A,\,
    K'(K'-1)/2-L_A
    \right).
\end{equation}
At $n_A=K'/2$, this becomes a reflection within the same sector. Fig.~\ref{fig:gallery}(f,g) shows exactly this structure: Pfaffian counting at one angular-momentum edge is mirrored by anti-Pfaffian counting at the other. Their simultaneous appearance follows from particle--hole symmetry and the resulting degeneracy between Pfaffian and anti-Pfaffian states at half-filled LL1 with Coulomb interaction~\cite{Levin2007APf,Lee2007APf}. The bootstrap therefore resolves Moore--Read-type local structure without selecting between the two particle--hole conjugates.

At LL1 $\nu=2/5$, the leading candidates for the $\nu=12/5$ ground state are Jain~\cite{Jain1989,Haldane1983}, Bonderson--Slingerland (BS)~\cite{BondersonSlingerland2008,Bonderson2012} and $\overline{\mathrm{RR}}_3$~\cite{ReadRezayi1999}, with previous numerical studies favoring $\overline{\mathrm{RR}}_3$ for the Coulomb interaction~\cite{RezayiRead2009,Zhu2015twelvefifths,Mong2017}. The bootstrap OES lacks the branch structure expected for the BS order, but is compatible with Jain counting $2,3,6$ from the lower-$L_A$ side and $\overline{\mathrm{RR}}_3$ counting $1,1,3,5$ from higher-$L_A$ side (Fig.~\ref{fig:gallery}(h)). Fidelity helps distinguish these two possibilities: the substantially lower Jain fidelity favors $\overline{\mathrm{RR}}_3$. Conditional orbital polarization supports the same conclusion, placing the bootstrap result closer to the $\overline{\mathrm{RR}}_3$ model wavefunction than to Jain or BS (see Extended Data Fig.~\ref{fig:twofifths} and its caption for more details). Together, these diagnostics point to $\overline{\mathrm{RR}}_3$ as the most likely Coulomb ground-state order at $\nu=12/5$.

Finally, at LLL $\nu=1/2$, the OES in Fig.~\ref{fig:gallery}(c) shows no isolated low-lying chiral branch, consistent with previous studies of the composite Fermi liquid~\cite{Geraedts2016DiracCF}. The quarter-filled LLL is particularly puzzling because incompressible states near $\nu=1/4$ have been reported in GaAs systems and bilayer graphene~\cite{Luhman2008,Wang2023Quarter,Kumar2025Quarter}, while theory admits competing possibilities ranging from a compressible four-flux CFL to several paired incompressible phases, including Pfaffian-like and other paired composite-fermion states~\cite{Wang2019DiracHierarchy,PapicMollerMilovanovicRegnaultGoerbig2009,FaugnoBalramBarkeshliJain2019,SharmaBalramJain2024,YutushuiMross2025}. The $\nu=1/4$ OES in Fig.~\ref{fig:gallery}(d) does not develop the isolated Pfaffian-like branch seen for the Moore--Read parent Hamiltonian or for the Coulomb ground state at $\nu=5/2$. This disfavors simple Pfaffian order, but the absence of an isolated branch does not by itself identify a CFL. We therefore turn to real-space pair correlations, whose oscillatory structure probes the composite Fermi surface.

\section*{Ground-state correlations in real space}

The entanglement spectrum identifies how the ground state is organized in Hilbert space. The same bootstrap also reveals how its particles organize in real space. We characterize this structure through the pair distribution function
\begin{equation}
    g(r)
    =
    \langle:\rho(\mathbf r)\rho(\mathbf 0):\rangle/n^2,
\end{equation}
where $n=\nu/(2\pi\ell_B^2)$ is the mean electron density. The pair distribution directly probes the correlation hole, screening and longer-range structure of quantum Hall fluids~\cite{Laughlin1983,GMP1986,RezayiRead1994,KamillaJainGirvin1997}. A window of $K$ symmetric-gauge orbitals resolves a characteristic radial scale $R_W\simeq\sqrt{2K}\ell_B$.

\begin{figure}[!tbp]
\centering
\includegraphics[width=0.9\columnwidth]{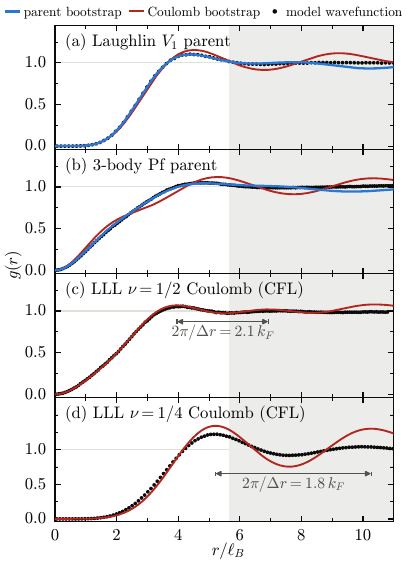}
\caption{\textbf{Real-space pair correlations from bootstrap.}
Pair distribution function $g(r)$ reconstructed from $\rho_W^\star$.
(a,b) For the Laughlin $V_1$ and three-body Moore--Read parent Hamiltonians, the bootstrap correlations agree almost point by point with Monte Carlo evaluations of the corresponding $N_e=200$ and $N_e=100$ model wavefunctions. The dashed curves show the Coulomb states at LLL $\nu=1/3$ and LL1 $\nu=1/2$, which preserve the correlation hole while developing stronger oscillatory structure.
(c,d) At LLL $\nu=1/2$ and $\nu=1/4$, the Coulomb states are compared with Jain composite-fermion model wavefunctions and exhibit longer-range Friedel oscillations. Their spacings give $2\pi/\Delta r\simeq2.1k_F$ and $1.8k_F$, respectively, close to the $2k_F$ scale of a composite Fermi surface. Shading marks distances beyond the window radius $R_W\simeq\sqrt{2K}\ell_B$ with $K=16$.}
\label{fig:gr}
\end{figure}

For the parent Hamiltonians, the bootstrap reproduces microscopic correlations as accurately as the entanglement spectrum. Fig.~\ref{fig:gr}(a,b) compares the bootstrap $g(r)$ with Laughlin and Moore--Read model wavefunctions~\cite{Laughlin1983,MooreRead1991}. The agreement is almost point by point throughout the resolved region, from the correlation hole through the subsequent oscillations. The short-distance behavior provides a sharp signature of the underlying clustering: states with a nonzero $m=1$ pair amplitude have $g(r)\propto r^2$, while the Laughlin $V_1$ parent removes this channel exactly and therefore gives $g(r)\propto r^6$~\cite{Haldane1983,TrugmanKivelson1985,Laughlin1983}. The Moore--Read parent and the Coulomb ground states retain the $m=1$ channel and show the corresponding $r^2$ behavior (Extended Data Fig.~\ref{fig:gr_loglog})~\cite{MooreRead1991}. At intermediate distances, the Coulomb correlations develop stronger oscillations, consistent with previous finite-system calculations at both LLL $\nu=1/3$ and half filling of LL1~\cite{BalramWojs2020}.

For the gapless states, the real-space correlations provide a more direct diagnostic than entanglement counting. At LLL $\nu=1/2$, the bootstrap $g(r)$ closely follows the Jain composite-fermion model wavefunction and develops the slowly damped oscillations characteristic of a composite Fermi liquid~\cite{RezayiRead1994,KamillaJainGirvin1997}. Their spacing gives
\begin{equation}
    \frac{2\pi}{\Delta r}\simeq2.1\,k_F ,
\end{equation}
with $k_F=\sqrt{4\pi n}=1/\ell_B$ at $\nu=1/2$, close to the expected $2k_F$ scale. Infinite-cylinder DMRG likewise finds singular density correlations associated with scattering across the composite-fermion Fermi surface~\cite{Geraedts2016DiracCF}.\footnote{The density oscillation observed here is compatible with the Dirac-composite-fermion description because suppression of exact $2k_F$ backscattering applies to the particle--hole-even channel.}

The same diagnostic resolves the ambiguity left by the OES at LLL $\nu=1/4$. The bootstrap \(g(r)\) exhibits Friedel oscillations with extrema at nearly the same positions as the Jain four-flux composite-fermion Fermi-sea wavefunction, although their amplitudes differ. The oscillation period gives \(2\pi/\Delta r\simeq1.8k_F\) in both cases, close to the expected \(2k_F\) scale of a composite Fermi surface. This behavior is also consistent with our independent DMRG study, which finds a composite Fermi liquid at $\nu=1/4$~\cite{WangFengZaletel2026}. Together with the absence of an isolated Pfaffian-like branch in the OES, these results identify the LLL $\nu=1/4$ Coulomb ground state as a composite Fermi liquid rather than a paired incompressible state. Thus the same optimized window density matrix provides complementary phase diagnostics: the OES constrains possible topological orders, while \(g(r)\) resolves the composite-Fermi-liquid structure.

\section*{Discussion}

Taken together, these results show that the proposed bootstrap scheme is a ground-state solver directly in the thermodynamic limit. The energy provides a global and systematically convergent measure of accuracy, fidelity directly benchmarks the window density matrix, and the OES and real-space correlations resolve the structure relevant to each phase. Applied to FQH systems, this combination identifies a composite Fermi liquid at LLL $\nu=1/4$ and suggests $\overline{\mathrm{RR}}_3$ as the most likely order at $\nu=12/5$.

The price of working directly in the thermodynamic limit is finite correlation range rather than finite system size. ED, DMRG and other finite-system methods are limited by the largest accessible particle or orbital number and approach the TDL by increasing that size. Here the particle number is infinite from the outset. Increasing $K$ instead extends the correlation range $R\sim\sqrt{K}\ell_B$ and systematically tightens both the energy bound and the information contained in the window density matrix.

The central observation is that this convergence extends beyond the energy. Fidelity provides a direct test of the density matrix itself: for the Laughlin and Moore--Read parents, $F$ remains $0.996$ and $0.948$ on a $K'=12$ cap, while established Coulomb states have high fidelity with their expected model states. The same $\rho_W^\star$ also reproduces detailed entanglement spectra and pair correlations. Together with the tightening energy bound, these results support
\begin{equation}
    \mathcal E^\star \simeq \mathcal E_{\rm gs},
    \qquad
    \rho_W^\star \simeq \rho_{{\rm gs},W}.
\end{equation}
Thus $\rho_W^\star$ is not merely an auxiliary variable certifying the energy, but an accurate representation of the thermodynamic ground state within the resolved window. In this sense, the bootstrap functions as a powerful ground-state solver directly in the thermodynamic limit.

At $\nu=12/5$, the close thermodynamic energies of Jain, BS and $\overline{\mathrm{RR}}_3$ make the competition physically consequential. Although our local diagnostics point most strongly to $\overline{\mathrm{RR}}_3$ for the ideal Coulomb interaction, small deviations from this interaction may change the relative energies and determine which phase is realized experimentally. The competition itself is therefore an important outcome, alongside the identification of the most likely Coulomb ground state.

This near-degeneracy also clarifies how to interpret the bootstrap density matrix. The bootstrap approaches the ground-state energy from below without restricting the optimization to a chosen trial-state family. When competing states have nearly equal energies, the energy objective may not distinguish their local correlations at the available resolution, and $\rho_W^\star$ can retain signatures associated with several candidates. The exact ground-state density matrix remains in the feasible set because it satisfies all imposed necessary constraints. The limitation is resolving it uniquely among close competitors. At particle--hole-symmetric half filling, the Pfaffian and anti-Pfaffian provide an example of simultaneous competing-order signatures. Resolving a small energy splitting requires sufficiently tight bounds and sufficiently informative local constraints.

\clearpage
\section*{Methods}

\subsection*{Thermodynamic-limit bootstrap}
\label{sec:tdl_bootstrap}
Different from the discussions in the main text, here we detail the constructions of the constraints. Some new notations are needed and properly defined.
The bootstrap is formulated directly in the thermodynamic limit (TDL), rather than on a finite system followed by extrapolation. On the Haldane sphere, the $n$th Landau level contains $2l+1$ orbitals, with $l=Q+n$ and $Q$ the monopole strength. We first take
\begin{equation}
    Q,N\rightarrow\infty,\qquad \frac{N}{2l+1}\rightarrow\nu,
\label{eq:tdl_limit}
\end{equation}
at fixed filling $\nu$. The orbitals near the north pole then approach the symmetric-gauge basis of the infinite plane. We retain a polar window $W=\{0,\ldots,K-1\}$ of $K$ orbitals, with all matrix elements and constraints evaluated directly in this limit. Thus $K$ controls the spatial extent of the resolved correlations, not the size of a finite physical system. Increasing $K$ strengthens the local constraints on the same infinite fluid.

\paragraph*{Window density matrix and homogeneity.}
The primary variable is a density matrix $\rho_W$ acting on the full $2^K$-dimensional Fock space of the window. The window contains no fixed number of particles because it exchanges particles with the surrounding infinite fluid. Writing $c_a$ for the orbital annihilation operator and $n_a=c_a^\dagger c_a$, we impose
\begin{equation}
    \rho_W\succeq 0,\qquad
    \operatorname{tr}\rho_W=1,\qquad
    \operatorname{tr}(\rho_W n_a)=\nu,
    \quad a\in W.
\label{eq:bootstrap_density}
\end{equation}
Particle-number conservation and axial rotational symmetry further organize $\rho_W$ into blocks labeled by $N_W=\sum_{a\in W}n_a$ and $L_W=\sum_{a\in W}a n_a$.

Homogeneity requires the correlations inside the window to be restrictions of common bulk relative-state correlations. For the two-particle reduced correlation matrix
$\Gamma^{(2)}_{ab,cd}=\operatorname{tr}(\rho_W c_a^\dagger c_b^\dagger c_d c_c)$, we impose
\begin{equation}
    \Gamma^{(2)}
    =\nu\sum_{\substack{m\le 2K-3\\m\ {\rm odd}}}
    \pi_m\Pi^{\mathrm{pl}}_{m,K},
    \qquad \pi_m\ge0.
\label{eq:bootstrap_pair}
\end{equation}
Here $\Pi^{\mathrm{pl}}_{m,K}$ is the planar projector onto relative angular momentum $m$, restricted to the window, and $\pi_m$ is the corresponding bulk pair weight, normalized such that $\pi_m=1$ for a filled Landau level.

For $k=3,4$, homogeneity takes the form
\begin{equation}
    \Gamma_T^{(k)}
    =\sum_{D\le T}
    C_{D,T}^{(k)}\gamma_D^{(k)}C_{D,T}^{(k)\dagger},
    \qquad
    \gamma_D^{(k)}\succeq0.
\label{eq:bootstrap_cluster}
\end{equation}
Here $\Gamma_T^{(k)}$ is the $k$-particle correlation matrix in the sector of total orbital-index sum $T$, with elements $\operatorname{tr}(\rho_W c_X^\dagger c_{X'})$ for ordered $k$-orbital configurations $X$ and $X'$. Here $c_X=c_{a_n}\cdots c_{a_1}$ with $a_i \in W$. The matrices $C_{D,T}^{(k)}$ contain center-of-mass descendants of relative-cluster states of degree $D$, restricted to the window. The positive matrices $\gamma_D^{(k)}$ encode the bulk relative-cluster weights and are independent of $T$. This common set of relative correlations across all center-of-mass sectors is the higher-body homogeneity constraint.

The same representation treats genuine higher-body interactions. For a two-body Hamiltonian, the energy is determined by the pair weights $\pi_m$. For a three-body Hamiltonian, the corresponding variables are the three-body cluster weights $\gamma_D^{(3)}$, and a general three-body contribution is
\begin{equation}
    \mathcal{E}_3[\gamma^{(3)}]
    =\sum_D\operatorname{tr}\!\left[
    \mathcal V_D^{(3)}\gamma_D^{(3)}
    \right],
\label{eq:bootstrap_threebody_energy}
\end{equation}
where $\mathcal V_D^{(3)}$ is the interaction matrix in the relative-cluster sector $D$, including the normalization appropriate to the energy per particle. The three- and four-body homogeneity constraints are also imposed for two-body Hamiltonians, where they strengthen local consistency without changing the physical interaction.

\paragraph*{Density fluctuations and energy optimization.}
For two-body interactions, we additionally impose positivity of guiding-center density fluctuations in the infinite fluid. Defining
\begin{equation}
    \rho_{\mathbf q}=\sum_j e^{-i\mathbf q\cdot\mathbf R_j},
    \qquad
    \delta\rho_{\mathbf q}=\rho_{\mathbf q}-\langle\rho_{\mathbf q}\rangle,
\end{equation}
the thermodynamic static structure factor satisfies
\begin{equation}
    S(\mathbf q)
    \equiv
    \lim_{N\to\infty}
    \frac{1}{N}
    \left\langle
    \delta\rho_{-\mathbf q}\delta\rho_{\mathbf q}
    \right\rangle
    \ge0.
\label{eq:bootstrap_structure_factor}
\end{equation}
For a rotationally invariant fluid,
\begin{equation}
    S(q)=1-\nu+4e^{-t/2}
    \sum_{m\,{\rm odd}}(\pi_m-\nu)L_m(t),
    \qquad t=q^2\ell_B^2,
\end{equation}
where $L_m$ is a Laguerre polynomial. A finite window fixes only the pair channels $m\le2K-3$. To impose $S(q)\ge0$ without truncating the infinite expansion, we introduce a non-negative measure $d\mu(t)\ge0$, corresponding to $d\mu(t)=S(\sqrt{t}/\ell_B)dt/2$ for a regular structure factor, and constrain its Laguerre moments,
\begin{equation}
    \int_0^\infty
    L_m(t)e^{-t/2}\,d\mu(t)
    =(-1)^m-\nu+2\pi_m,
    \qquad 0\le m\le\bar m.
\label{eq:bootstrap_moments}
\end{equation}
Here $\pi_m=0$ for even $m$, and unresolved odd-channel weights remain non-negative auxiliary variables. We take $\bar m=2K-3$, so the thermodynamic completion matches exactly the pair channels resolved by the window. Numerically, measure positivity is enforced by non-negative grid weights together with a polynomial cone for the high-momentum tail. This constrains density fluctuations at all momenta without introducing a real-space interaction cutoff.

The neutral two-body interaction energy is the linear functional
\begin{align}
    \mathcal{E}_2[\mu]
    &=\frac12\int_0^\infty e^{-t/2}f(t)\,d\mu(t)
    -\frac12 w_V(0),
    \label{eq:bootstrap_energy}\\
    f(t)
    &=\frac{V(\sqrt t)}{2\pi}
    \left[L_n(t/2)\right]^2,
\label{eq:bootstrap_kernel}
\end{align}
where $\ell_B=1$, $V(q)$ is the Fourier-transformed interaction, $n$ labels the Landau level, and $w_V(0)$ is the known self-energy contribution. For Coulomb interactions, $V(q)=2\pi/q$. All energies are quoted in units of $e^2/(\epsilon\ell_B)$.

The primal SDP minimizes $\mathcal{E}_2[\mu]$ over $\rho_W$, $\pi_m$, $\gamma_D^{(k)}$ and $\mu$, subject to Eqs.~(\ref{eq:bootstrap_density})--(\ref{eq:bootstrap_moments}). The corresponding three-body program uses Eq.~(\ref{eq:bootstrap_threebody_energy}) as its objective. Hamiltonians containing both interactions simply add the two linear contributions. In particular, the Coulomb objective retains the full interaction at all momenta.

% \paragraph*{Direct thermodynamic-limit certification.}
% The rigorous energy statement follows from the dual SDP and does not assume that the optimized window density matrix extends to a physical infinite-system state. For the two-body Coulomb problem, the dual combines positive window and relative-cluster operators with pair-channel consistency conditions and the continuous density-kernel inequality
% \begin{equation}
%     2P(t)\le f(t),
%     \qquad
%     P(t)=\sum_{m\le\bar m}a_m L_m(t),
%     \qquad t\ge0.
% \label{eq:bootstrap_dual_kernel}
% \end{equation}
% A verified positive window operator can be embedded in arbitrarily large spheres and averaged over rotations, converting the local operator inequality into an extensive lower bound. Combining this inequality with the pair-channel and density-fluctuation constraints gives
% \begin{equation}
%     H_l-\frac{N^2U_0}{2}
%     \ge N\mathcal{E}^\star+o(N),
%     \qquad
%     \mathcal{E}_0(\nu)\ge\mathcal{E}^\star,
% \label{eq:bootstrap_tdl_bound}
% \end{equation}
% where $U_0$ is the monopole Hartree coefficient and $\mathcal{E}_0(\nu)$ is the neutral ground-state energy per particle in the thermodynamic limit. The $o(N)$ term vanishes relative to $N$ as $Q,N\rightarrow\infty$ at fixed filling. Every accepted certificate therefore directly bounds the TDL ground-state energy without extrapolation in particle number or window size.

\paragraph*{Direct thermodynamic-limit certification.}
The rigorous energy statement follows from the dual SDP and does not assume that the optimized window density
matrix extends to a physical infinite-system state. The multipliers of the window, pair and cluster constraints
($\tau$, $\lambda_a$, $\Lambda$, $Y^{(k)}$) and of the structure-factor moments ($a_m$) define a window operator
and a dual objective,
\begin{equation}
\begin{aligned}
    \mathcal Z_W&=L_2(\Lambda)-\tau-\sum_{a<K}\lambda_an_a-\sum_{k=3,4}N_k,\\
    \mathcal{E}^\star&=\frac{\tau}{\nu}+\sum_{a<K}\lambda_a+\sum_{m\le\bar m}a_m\big[(-1)^m-\nu\big]-\frac12w_V(0),
\end{aligned}
\label{eq:bootstrap_dual_objective}
\end{equation}
where $L_2(\Lambda)=\sum\Lambda_{ab,cd}\,c_a^\dagger c_b^\dagger c_dc_c$ and $N_k$ is the second quantization of
$Y^{(k)}$. Feasibility requires $\mathcal Z_W\succeq0$ in every $(N_W,L_W)$ block, positive semidefinite cluster
Gram matrices $\mathcal M^{(k)}_D$, pair budgets $2a_m\ge\operatorname{tr}(\Lambda\Pi^{\rm pl}_{m,K})$ for odd
$m\le2K-3$, and
\begin{equation}
    2P(t)\le f(t),
    \qquad
    P(t)=\sum_{m\le\bar m}a_m L_m(t),
    \qquad t\ge0.
\label{eq:bootstrap_dual_kernel}
\end{equation}
The primal energy then exceeds $\mathcal{E}^\star$ by a sum of nonnegative terms. The last of these is
$\frac12\int e^{-t/2}[f(t)-2P(t)]\,d\mu(t)$ over the structure-factor measure, which carries the unresolved pair
channels and the long-range Coulomb tail. Any feasible dual point is therefore a lower bound, optimal or not, and
the finite window imposes no cutoff on the interaction.

An independent verifier turns the floating-point dual into a certificate. It rounds the multipliers to
rationals, proves positivity by exact integer elimination (a rigorous Cholesky test for the largest window
blocks), and checks the budgets in interval arithmetic. The solver imposes Eq.~\eqref{eq:bootstrap_dual_kernel}
only on a momentum grid, so the $a_m$ are re-optimized by a small linear program and the kernel inequality is then
proved on the whole half-line. For Coulomb this uses exact Sturm root counting and negative tail coefficients.
The certified $\mathcal{E}^\star$, rounded down at $10^{-7}$, lies within $2\times10^{-6}$ of the floating-point optimum.

Embedding $\mathcal Z_W$ in an arbitrarily large sphere and averaging over rotations preserves positivity and
makes the local terms extensive~\cite{OpenAI2026Fock}. The strict margins of the verified conditions absorb the
$O(1/l)$ curvature corrections, and density-fluctuation positivity controls the rest of the interaction. For
$Q\ge Q_0$, with $Q_0$ existential,
\begin{equation}
    H_l-\frac{N^2U_0}{2}
    \ge N\mathcal{E}^\star+o(N),
    \qquad
    \mathcal{E}_0(\nu)\ge\mathcal{E}^\star,
\label{eq:bootstrap_tdl_bound}
\end{equation}
where $U_0$ is the monopole Hartree coefficient, $\mathcal{E}_0(\nu)$ is the neutral ground-state energy per
particle in the thermodynamic limit, and $o(N)/N\to0$ as $Q,N\to\infty$ at fixed filling. The bound is state
independent, so it covers symmetry-broken states, and it needs no extrapolation in particle number or window
size.

The numerical SDP represents the density-fluctuation measure by $4{,}040$ grid points up to $t=50$, together with a polynomial tail cone. After optimization, the dual is independently verified, including positivity of the window and cluster operators and the continuous kernel inequality between grid points and throughout the tail. The extension of this certification procedure to genuine three-body Hamiltonians uses the corresponding three-cluster dual conditions.

\subsection*{Ground-state diagnostics}

\paragraph*{Orbital entanglement spectrum.}
The optimized window state $\rho_W^\star$ acts on the full Fock space of the $K$ window orbitals. To obtain the orbital entanglement spectrum (OES), we choose an inner cap
$A=\{0,\ldots,K'-1\}$ and trace out the remaining window orbitals,
\begin{equation}
    \rho_A
    =
    \operatorname{tr}_{K',\ldots,K-1}\rho_W^\star .
\end{equation}
The cap conserves particle number and orbital angular momentum,
\begin{equation}
    \hat n_A=\sum_{a=0}^{K'-1}n_a,
    \qquad
    \hat L_A=\sum_{a=0}^{K'-1}a\,n_a .
\end{equation}
We diagonalize $\rho_A$ in each $(n_A,L_A)$ block. Its eigenvalues
$\lambda$ define the entanglement energies
$\xi=-\ln\lambda$~\cite{LiHaldane2008,Thomale2010}.

The finite cap has its own edge counting and need not display the
asymptotic thermodynamic sequence at all angular momenta. We therefore
compare with the counting appropriate to the same $(K',n_A)$.
For the Laughlin state, this is the finitized chiral-boson counting,
obtained by counting the allowed electron displacements at fixed
$\Delta L$ before an electron reaches the end of the cap. For
$K'=12$ and $n_A=4$, for example, the resulting sequence is
$1,1,2,3,5,6,8,\ldots$, rather than the unrestricted thermodynamic
partition sequence at large $\Delta L$~\cite{Thomale2010,
ChandranHermannsRegnaultBernevig2011}.
For the Moore--Read state we enumerate the corresponding finite-cap
$(2,4)$-admissible configurations, in which no more than two electrons
occupy four consecutive orbitals. This gives
$1,2,4,6,9,11,\ldots$ for $(K',n_A)=(10,5)$ and
$1,1,3,5,9,12,18,\ldots$ for $(12,6)$.
For $\overline{\mathrm{RR}}_3$, we analogously enumerate the
particle--hole conjugate of the $(3,5)$-admissible Read--Rezayi
configurations. For the $(K',n_A)=(10,4)$ sector used in
Fig.~\ref{fig:gallery}, the branch begins with $1,1,3,5$ when read from
the higher-$L_A$ hole edge. The Jain reference counting is the rank, in
each $(n_A,L_A)$ sector, of the span of composite-fermion droplet states
in two $\Lambda$ levels, with Jain--Kamilla projection and restriction
to the $K'$ cap orbitals~\cite{JainKamillaPRB1997,RodriguezSimonSlingerland2012,DavenportRodriguezSlingerlandSimon2015}.
For the $(K',n_A)=(10,4)$ cap, this gives $2,3,6$ from the lower-$L_A$
side, as indicated in Fig.~\ref{fig:gallery}(h). The BS edge contains a forward
Ising$\times U(1)$ sector and a counter-propagating $U(1)$ sector, but
there is no Jack root-state construction that supplies a unique
finitized counting for the small cap used here~\cite{BondersonSlingerland2008}.
We therefore do not assign a BS counting sequence in Fig.~\ref{fig:gallery}(h).

\paragraph*{Particle entanglement spectrum.}
The particle entanglement spectrum (PES) is obtained directly from the
$n$-particle reduced density matrix of $\rho_W^\star$. We use $n$ here
to distinguish the particle cut from the fluctuating cap occupation
$n_A$ above. For ordered orbital configurations
$X=(a_1<\cdots<a_n)$, define
\begin{equation}
    [\rho_{\rm P}^{(n)}]_{X,X'}
    =
    \frac{
    \operatorname{tr}
    \!\left(
    \rho_W^\star c_{X'}^\dagger c_X
    \right)}
    {
    \left\langle
    \binom{\hat N_W}{n}
    \right\rangle
    }.
\end{equation}
The matrix is
resolved by total orbital index
$L=\sum_i a_i$, and its eigenvalues define
$\xi=-\ln\lambda$~\cite{SterdyniakRegnaultBernevig2011,
ChandranHermannsRegnaultBernevig2011}.

For model FQH states, the universal low-lying PES counting is governed by the corresponding quasihole Hilbert space~\cite{SterdyniakRegnaultBernevig2011,ChandranHermannsRegnaultBernevig2011}. For the Laughlin $(1,m)$ sequence, the quasihole space of $n$ particles with $N_{\rm qh}$ excess flux quanta has dimension
\begin{equation}
    \mathcal N_{\rm qh}
    =
    \binom{n+N_{\rm qh}}{n}.
\end{equation}
Its angular-momentum decomposition follows from the $(1,m)$-admissible configurations. The Moore--Read and Read--Rezayi sectors follow analogously from the $(2,4)$ and $(3,5)$ admissibility rules, respectively~\cite{BernevigHaldane2008Jack,BernevigRegnault2009}. For the Jain $2/5$ comparison, the target ranks are instead obtained from lowest-Landau-level-projected composite-fermion droplet states~\cite{RodriguezSimonSlingerland2012,DavenportRodriguezSlingerlandSimon2015}. For particle--hole-conjugate states we construct the corresponding hole RDM by exchanging particles and holes. In the bootstrap there is no finite total particle number or flux: $\rho_{\rm P}^{(n)}$ is formed directly from the TDL window state. We use only sectors with $L\leq K-1$, for which every contributing $n$-particle configuration lies entirely inside the window and the block is therefore unaffected by the orbital boundary.

\paragraph*{Conditional orbital polarization.}
The block weights of $\rho_A$ also measure how particles at fixed
$n_A$ are distributed across the cap. We define
\begin{equation}
    \Delta
    =
    \frac{
    \displaystyle\sum_{L_A}
    \left[
    L_A-\frac{n_A(K'-1)}{2}
    \right]
    \operatorname{tr}\rho_A^{(n_A,L_A)}
    }{
    \displaystyle\sum_{L_A}
    \operatorname{tr}\rho_A^{(n_A,L_A)}
    },
    \qquad
    n_A=\nu K',
\end{equation}
whenever $\nu K'$ is an integer. The subtraction removes the angular
momentum of a uniform distribution of $n_A$ particles across the cap.
Particle--hole conjugation sends $\Delta\rightarrow-\Delta$.
We use this quantity at LL1 $\nu=2/5$ as a local diagnostic
complementary to the OES, with the LLL Jain state at the same filling
as a control.

For a rotationally invariant finite-sphere reference, the filling is $\nu_N=N_e/(2l+1)$, so the mean cap occupation $\bar n=\nu_N K'$ generally differs from the integer $\nu K'$ used for the bootstrap. We define $d(n)=\langle L_A\rangle_{n_A=n}-n(K'-1)/2$ and fit
\begin{equation}
    d(n)=a+b(n-\bar n)+c(n-\bar n)^2
\end{equation}
with weights $P(n)=\operatorname{tr}\rho_A^{(n)}$. The uniform-state sum rule $\sum_nP(n)d(n)=0$ fixes $a=-c\,\operatorname{Var}(n_A)$, and we report
\begin{equation}
    \widehat\Delta=-c\,\operatorname{Var}(n_A).
\end{equation}
This estimator accounts for the finite-sphere filling offset but does not remove curvature corrections. Tests on exact Laughlin spheres with $N_e=8$--$14$ reproduce the thermodynamic six-orbital-cap value to $0.004$. On the ten-orbital cap used for phase identification, the finite-sphere estimate differs from the thermodynamic value by $0.04$--$0.06$. This calibration indicates the scale of finite-size effects, rather than a statistical error bar for each candidate.

\paragraph*{Pair correlation function.}
The two-particle sector of the bootstrap gives the pair weights
$\pi_m$ in relative-angular-momentum channels. These determine the
guiding-centre pair distribution function,
\begin{equation}
    g(r)
    =
    \frac{2}{\nu}
    \sum_{m\,{\rm odd}}
    \pi_m
    \frac{x^m e^{-x}}{m!},
    \qquad
    x=\frac{r^2}{4\ell_B^2}.
\label{eq:methods_gr}
\end{equation}
A $K$-orbital window determines the channels $m\leq2K-3$.
For reconstructing $g(r)$ we set unresolved higher channels to their
uncorrelated value $\pi_m=\nu$. Their contribution becomes relevant
only near and beyond the characteristic window radius
$R_W\simeq\sqrt{2K}\ell_B$. The shaded regions in
Fig.~\ref{fig:gr} mark this scale. The short-distance behavior is
controlled entirely by the lowest relative-angular-momentum channels
and is therefore insensitive to this completion.

\paragraph*{Window-size extrapolation.}
The first $K$ symmetric-gauge orbitals occupy an area of order $2\pi K\ell_B^2$, so the characteristic linear scale of the window is
\begin{equation}
    R_W\simeq\sqrt{2K}\,\ell_B.
\end{equation}
The residual dependence on $K$ therefore measures finite correlation range rather than finite physical size. In Extended Data Fig.~\ref{fig:energy_extrap}, gapped states are fit to corrections exponential in $R_W$, as expected for a finite correlation length, and gapless states are fit to algebraic corrections in $R_W$. These extrapolations are used only to characterize convergence and never enter the certification of a finite-$K$ bound.

\subsection*{Reference calculations}

The model references use different numerical representations because the tractable system size and the cost of obtaining a full cap density matrix, an entanglement spectrum or a pair correlation function differ substantially. We use conformal-field-theory matrix product states (CFT-MPS) for the accessible model cap density matrices and OES~\cite{ZaletelMong2012,EstiennePapicRegnaultBernevig2013}, Monte Carlo sampling for projected composite-fermion cap density matrices and large-system $g(r)$, and exact finite-sphere model states for conditional polarization. Disk bulk regions avoid sphere curvature and give smaller finite-size effects when a uniform interior is accessible. Some multicomponent disk constructions have nonuniform central occupations. For these model wavefunctions we use magnetically translated caps away from the center and edge, or rotationally invariant sphere states. Sphere model wavefunctions retain curvature corrections of order $1/N_e$. The geometry, construction and size are specified below for each comparison. These finite-size effects belong to the model wavefunctions, whereas the bootstrap is formulated directly in the TDL.

\paragraph*{Thermodynamic energy benchmarks.}
Finite-sphere exact diagonalization uses the Coulomb pseudopotentials of the corresponding Landau level and the relevant shift and parity branches for each filling. We calculate systems with $N_e=6$--$18$ electrons, depending on the problem, with the largest calculations containing up to $38$ active Landau-level orbitals. A uniform neutralizing background with surface charge density $\sigma_{\rm b}=N_e e/(4\pi R^2)$ is included on a sphere of radius $R=\sqrt{N_\phi/2}\,\ell_B$. If $E_{\rm ee}$ denotes the electron--electron interaction energy, the density-corrected neutral energy per particle is
\begin{equation}
    \mathcal E_{N_e}
    =
    \sqrt{\frac{\nu N_\phi}{N_e}}
    \left(
    \frac{E_{\rm ee}}{N_e}
    -
    \frac{N_e}{2R}
    \right).
\end{equation}
The factor $\sqrt{\nu N_\phi/N_e}$ corrects the finite-sphere density to the target thermodynamic density~\cite{Morf1986,BalramTokeWojsJain2015}. The ED energies are extrapolated linearly in $1/N_e$. At LL1 $\nu=2/5$ we additionally calculate the Bonderson--Slingerland shift $\mathcal S=2$ over $N_e=6$--$14$ where available; the branch shown in Fig.~\ref{fig:schematic}(a) and Extended Data Fig.~\ref{fig:ed_extrap}(f) uses $N_e=8$--$14$, all with $L=0$. The uncertainty shown in Fig.~\ref{fig:energy} is the larger of the leave-one-out variation of the extrapolated value and, when available, the difference obtained from an alternative shift or parity branch, and therefore measures sensitivity to the finite-size extrapolation rather than statistical uncertainty. Published DMRG thermodynamic estimates are taken from Refs.~\cite{ZhaoShengHaldane2011,ZaletelMongPollmannRezayi2015}.

\paragraph*{Entanglement-spectrum benchmarks.}
The model spectra in Fig.~\ref{fig:gallery} are obtained by contracting disk CFT-MPS representations of the Laughlin and Moore--Read states~\cite{Laughlin1983,MooreRead1991,ZaletelMong2012,EstiennePapicRegnaultBernevig2013}. We use $N_e=40$ electrons with CFT descendant-level cutoff $P=24$ for Laughlin and $P=20$ for Moore--Read. The cap is centred at the origin, within the uniform interior of these droplets. The same $K'$-orbital cut is applied to the model wavefunctions and to $\rho_W^\star$, so the resulting OES can be compared level by level. The LLL $\nu=1/3$ comparison in Fig.~\ref{fig:gallery}(b) uses an independent $N_e=24$ sphere-DMRG ground state.

\paragraph*{Cap fidelity.}
For each $K=16$ bootstrap state we form $\rho_A^\star=\operatorname{tr}_{W\setminus A}\rho_W^\star$ and compare it with the corresponding model-state cap density matrix using $F(\rho,\sigma)=[\operatorname{Tr}\sqrt{\sqrt{\rho}\sigma\sqrt{\rho}}]^2$~\cite{Jozsa1994}. The Laughlin and Pfaffian model wavefunctions are disk CFT-MPS states with $N_e=40$ and cutoffs $P=28$ and $24$, respectively; the anti-Pfaffian is obtained by particle--hole conjugation~\cite{Levin2007APf,Lee2007APf}. The $\overline{\mathrm{RR}}_3$ model wavefunction is the particle--hole image of a disk Read--Rezayi state with $N=42$ particles and $P=8$~\cite{ReadRezayi1999,RezayiRead2009}. Its central five-orbital cap agrees with the corresponding cap obtained from the $N=30$ sphere Read--Rezayi model wavefunction to $F=0.984$.

The Jain $2/5$ and BS fidelity model wavefunctions are Jain--Kamilla-projected disk states with $N_e=42$~\cite{Jain1989,JainKamillaPRB1997,BondersonSlingerland2008}. For BS we use
\begin{equation}
    \Psi_{\rm BS}
    =
    {\rm Pf}\!\left(\frac{1}{z_i-z_j}\right)
    \Phi_1\,
    \mathcal P_{\rm JK}\!\left[\overline{\Phi}_2\Phi_1^2\right],
\end{equation}
where $\Phi_j$ denotes $j$ filled Landau levels and the bar reverses the magnetic field. Their cap density matrices are estimated from Monte Carlo expectation values of normal-ordered Fock-space operators, using $8\times10^6$ samples per state. The cap orbitals are magnetically translated to a bulk point away from both the centre and the edge, avoiding nonuniform central occupations. With the same estimator and statistics, the Laughlin state reproduces its exact five-orbital cap density matrix to $F=0.978$, setting an approximate fidelity resolution of $0.02$.

For both $\nu=1/2$ and $1/4$, the composite-Fermi-sea fidelity model wavefunctions use the Jain--Kamilla disk construction described below, with closed-shell sizes $N_e=37$ ($N_e=21$ as a size check) and the same translated bulk-cap estimator. The calculations use $3.4\times10^5$ and $2.4\times10^6$ samples, respectively, distributed over 12 random seeds at each size. Table~\ref{tab:fidelity} reports the $N_e=37$ results. On the five-orbital cap, the two sizes agree within $0.004$ at $\nu=1/4$ and within the Monte Carlo resolution of approximately $0.1$ at $\nu=1/2$.

Table~\ref{tab:fidelity} uses $K'=8$ for the parent, Laughlin and Pfaffian comparisons and a common $K'=5$ for the two-fifths and CFL comparisons, the largest cap available for the Monte Carlo density matrices. The larger parent benchmarks quoted in the text use $K'=12$. All fidelities are evaluated at the stated finite sizes without extrapolation. The $\nu = 12/5$ candidates are compared on the same cap size.

\paragraph*{Exact finite-sphere polarization benchmarks.}
The polarization benchmarks in Extended Data Fig.~\ref{fig:twofifths} are obtained from model states in the sphere Fock basis, with the same $K'$-orbital polar cap as the bootstrap. Laughlin calibration states are generated as fermionic Jack polynomials by the squeezing recursion from the root $100100\cdots$, for $N_e=6$--$14$ and $2l=3(N_e-1)$~\cite{BernevigHaldane2008Jack,BernevigRegnault2009}. The $\mathrm{RR}_3$ state is obtained by Lanczos diagonalization as the unique zero mode of the four-body projector $H_4=B^\dagger B$ in the subspace dominated by the $11100\,11100\cdots$ root. Here $B$ annihilates fermionic quartets with maximal angular momentum $L_4=4l-6$, with the magnetic components summed in $B^\dagger B$. At $2l=5N/3-3$ this is equivalently the fermionic $(k,r)=(3,2)$ Jack state~\cite{ReadRezayi1999,BernevigHaldane2008Jack}. Particle--hole conjugation gives the $\overline{\mathrm{RR}}_3$ model wavefunction. The sizes $N=12,15,18,21$ correspond to $N_e=6,8,10,12$ after conjugation.

The Jain $2/5$ model wavefunction is constructed by exact LLL projection of $\Phi_2\Phi_1^2$, with multiplication and projection performed in the Schur basis. We use $N_e=8,10$ at $2l=5N_e/2-4$. The BS model wavefunction is the exact product, evaluated in the Schur basis, of the bosonic $\nu=1$ Moore--Read state and the reverse-flux Jain $2/3$ state $\mathcal P_{\rm LLL}[\Phi_1^2\overline{\Phi}_2]$, for $N_e=6,8,10$ at $2l=5N_e/2-2$. This exact finite-sphere construction is distinct from the Jain--Kamilla disk implementation used for fidelity. The alternative projection prescription $[{\rm Pf}(1/(z_i-z_j))\Phi_1^2]\mathcal P_{\rm LLL}[\overline{\Phi}_2\Phi_1]$ gives $\widehat\Delta=-0.135$, compared with $-0.122$ for the first BS construction at $N_e=10$, $K'=10$. All sphere polarizations use the filling-offset estimator defined above.

\paragraph*{Pair-correlation benchmarks.}
For the model-state benchmarks, we use the Laughlin and Moore--Read wavefunctions on the disk,
\begin{align}
    \Psi_{\rm L}
    &=
    \prod_{i<j}(z_i-z_j)^3
    e^{-\sum_i |z_i|^2/4},\\
    \Psi_{\rm MR}
    &=
    {\rm Pf}\!\left(\frac{1}{z_i-z_j}\right)
    \prod_{i<j}(z_i-z_j)^2
    e^{-\sum_i |z_i|^2/4},
\end{align}
in symmetric gauge with $\ell_B=1$. Their pair correlations are evaluated by Metropolis sampling of $|\Psi|^2$, using $N_e=200$ electrons for the Laughlin state and $N_e=100$ for the Moore--Read state. To suppress edge effects, we accumulate pair-separation histograms using only electrons in the bulk of the droplet as the origins, and normalize by the corresponding bulk density to obtain $g(r)$.

For the composite Fermi liquids at $\nu=1/(2p)$, with $p=1$ for $\nu=1/2$ and $p=2$ for $\nu=1/4$, we use the Rezayi--Read composite-fermion Fermi-sea wavefunction~\cite{RezayiRead1994},
\begin{equation}
    \Psi_{\rm CFFS}
    =
    \mathcal P_{\rm LLL}
    \prod_{i<j}(z_i-z_j)^{2p}
    \det_{\alpha j} e^{i\mathbf k_\alpha\cdot\mathbf r_j}.
\end{equation}
Lowest-Landau-level projection is implemented using the Jain--Kamilla construction~\cite{JainKamillaPRB1997,KamillaJainGirvin1997}. The occupied momenta form closed-shell approximations to a circular composite Fermi sea with
\begin{equation}
    k_F=\frac{\sqrt{2\nu}}{\ell_B},
    \qquad
    d_k=k_F\sqrt{\frac{\pi}{N_e}},
\end{equation}
where $\mathbf k=d_k(m,n)$ lies on a square momentum lattice. No variational parameters are optimized, including the Fermi-surface shape. We use the closed-shell sizes $N_e=61$ and $97$.

The CFFS pair correlations are obtained by Metropolis sampling of $|\Psi_{\rm CFFS}|^2$ on the disk. We accumulate pair-separation histograms up to $0.55R$, using only electrons within $0.42R$ of the droplet centre as the origins, where $R=\sqrt{2N_e/\nu}$. Because Jain--Kamilla projection slightly modifies the droplet radius, $g(r)$ is normalized by the measured bulk density rather than the nominal value $\nu/(2\pi)$. The measured density is approximately $0.984$ of the nominal density at both fillings. Independent Monte Carlo runs give statistical errors of $8\times10^{-4}$--$1.6\times10^{-3}$ per bin, while the $N_e=61$ and $97$ results differ by at most $0.02$ over the comparison window. As consistency checks, the screening sum lies between $-0.99$ and $-1.00$. The Coulomb energies extracted from $g(r)$ are $-0.4665\,e^2/(\epsilon\ell_B)$ at $\nu=1/2$ for $N_e=61$ and $-0.3555\,e^2/(\epsilon\ell_B)$ at $\nu=1/4$ for $N_e=97$. The former agrees with established composite-Fermi-sea benchmarks~\cite{KamillaJainGirvin1997}.

\section*{Data and code availability}
The solver, verifier, and data will be made available upon reasonable request.

\section*{Author contributions}
Q.G. and T.W. contributed equally to the conceptualization of the project and the data analysis.

\section*{Acknowledgements}
The authors thank the MIT Office of Research Computing and Data (ORCD) for providing computational resources. The ED and DMRG benchmarks were performed using \texttt{Mosaic}, developed by T.W. The TDL bootstrap code, SDP solver and model-wavefunction Monte Carlo were constructed by the artificial-intelligence (AI) models Claude (Anthropic) and ChatGPT (OpenAI) under the authors' instructions. Q.G. acknowledges support from the National Science Foundation (NSF) Designing Materials to Revolutionize and Engineer our Future (DMREF) Grant No.~6955166. T.W. acknowledges support from the Harvard Quantum Initiative Fellowship and the Simons Collaboration on Ultra-Quantum Matter through Simons Foundation Grant No.~651440.

\bibliography{references}
\clearpage
% !TeX root = manuscript.tex
\onecolumngrid
\section*{Extended Data Figures}
\setcounter{figure}{0}
\renewcommand{\thefigure}{\arabic{figure}}
\renewcommand{\theHfigure}{ED.\arabic{figure}}
\makeatletter
\renewcommand{\fnum@figure}{Extended Data Fig.\ \thefigure}
\makeatother

\begin{figure}[!htbp]
\centering
\includegraphics[width=0.56\textwidth,height=0.56\textheight,keepaspectratio]{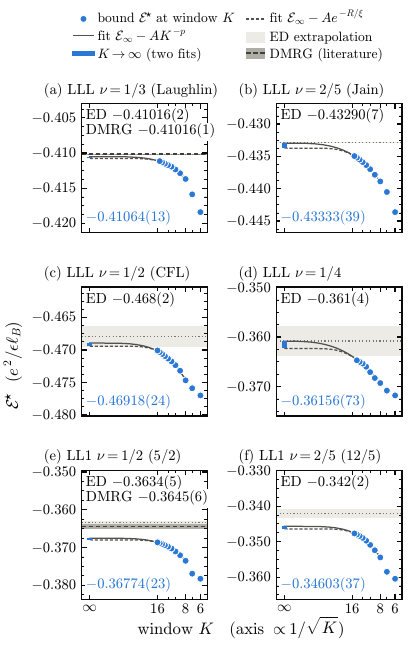}
\caption{\textbf{Finite-window convergence of the thermodynamic energy bounds.}
Certified bounds $\mathcal E^\star(K)$ for the six Coulomb problems in main-text Fig.~\ref{fig:energy}.
Because a window of $K$ Landau-level orbitals has area proportional to $K$, its characteristic linear extent scales as $R\propto\sqrt K\,\ell_B$; the horizontal axis is therefore proportional to $1/R$.
For a gapped state with a finite correlation length, finite-window corrections are naturally expected to decay exponentially in $R$, whereas gapless states can exhibit algebraic convergence.
We accordingly compare fits of the form
$\mathcal E_\infty-Ae^{-R/\xi}$ and
$\mathcal E_\infty-AK^{-p}$.
The resulting $K\rightarrow\infty$ estimates are compared with ED and DMRG thermodynamic values.
These extrapolations are used only to characterize the residual window dependence and do not enter the certified bounds in main-text Fig.~\ref{fig:energy}.}
\label{fig:energy_extrap}
\end{figure}
\clearpage

\begin{figure}[!htbp]
\centering
\includegraphics[width=0.72\textwidth,height=0.72\textheight,keepaspectratio]{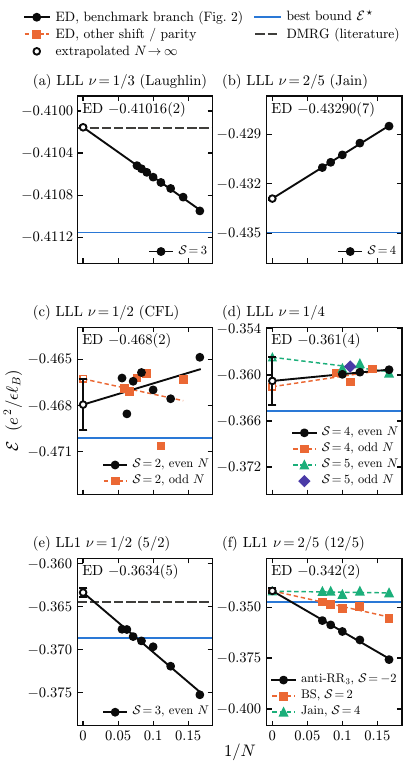}
\caption{\textbf{Finite-size extrapolation of the ED energy benchmarks.}
Finite-size Coulomb energies $\mathcal E$ versus $1/N$ for the six problems in main-text Fig.~\ref{fig:energy}. Black symbols show the benchmark shift and parity used for the thermodynamic extrapolation, while colored symbols show alternative finite-size branches where relevant. In panel (f), the LL1 $\nu=2/5$ alternatives include the $\overline{\mathrm{RR}}_3$ branch at $\mathcal S=-2$, the Bonderson--Slingerland branch at $\mathcal S=2$ and the Jain branch at $\mathcal S=4$. The extrapolated $N\rightarrow\infty$ values and their uncertainties define the ED bands shown in main-text Fig.~\ref{fig:energy}. The strongest certified bootstrap bound is shown in blue and DMRG estimates are included where available. The pronounced dependence on size, shift and parity in several panels illustrates the uncertainty inherent in extracting thermodynamic energies from finite systems.}
\label{fig:ed_extrap}
\end{figure}
\clearpage

\begin{figure}[!htbp]
\centering
\includegraphics[width=\textwidth]{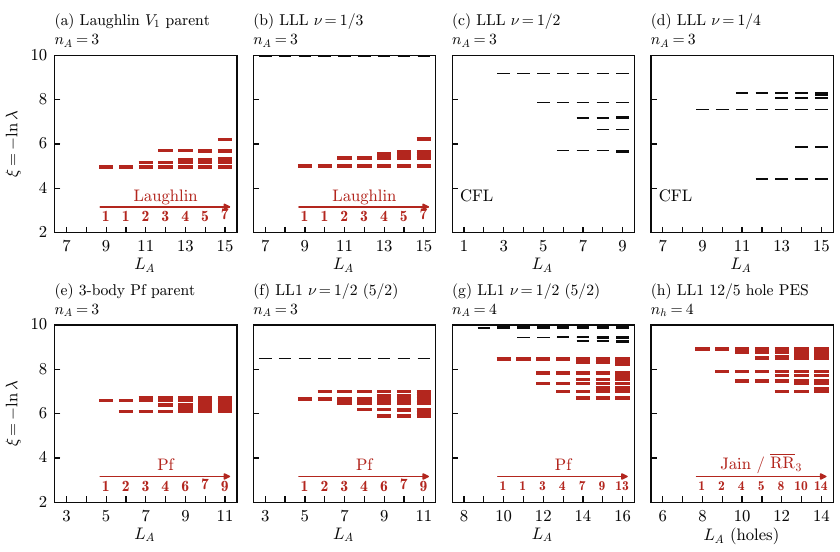}
\caption{\textbf{Particle entanglement spectra from bootstrap.}
Particle entanglement energies $\xi=-\ln\lambda$, resolved by particle number and angular momentum. 
(a,b) The Laughlin $V_1$ parent and LLL $\nu=1/3$ reproduce the Laughlin quasihole counting for $n_A=3$.
(c,d) LLL $\nu=1/2$ and $\nu=1/4$ show the corresponding composite-Fermi-liquid spectra.
(e,f) The three-body Moore--Read parent and LL1 $\nu=1/2$ reproduce the Pfaffian three-particle counting, while (g) shows the four-particle spectrum at LL1 $\nu=1/2$.
(h) The four-hole spectrum at LL1 $\nu=2/5$ has the same low-lying counting for the Jain and $\overline{\mathrm{RR}}_3$ candidates.
Red bars mark the low-lying levels entering the indicated counting.
The coincidence in (h) shows that the PES counting alone does not distinguish Jain from $\overline{\mathrm{RR}}_3$. The main-text OES compares the candidate edge structures, while the cap fidelity gives the direct state-level comparison~\cite{SterdyniakRegnaultBernevig2011,ChandranHermannsRegnaultBernevig2011}.}
\label{fig:pes}
\end{figure}
\clearpage

\begin{figure}[!htbp]
\centering
\includegraphics[width=0.48\textwidth,height=0.48\textheight,keepaspectratio]{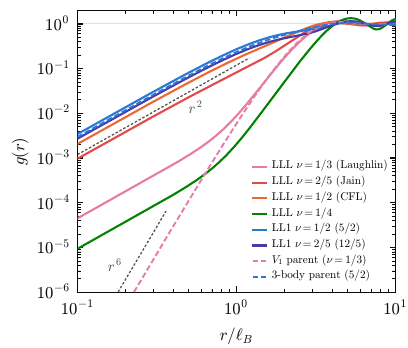}
\caption{\textbf{Short-distance scaling of the pair correlation function.}
Pair distribution functions $g(r)$ on logarithmic axes. A nonzero relative-angular-momentum-$m=1$ pair amplitude gives $g(r)\propto r^2$ as $r\rightarrow0$. The Coulomb states and the three-body Moore--Read parent show this quadratic behavior, whereas the Laughlin $V_1$ parent eliminates the $m=1$ channel and begins with the $m=3$ contribution, giving $g(r)\propto r^6$~\cite{Haldane1983,TrugmanKivelson1985,Laughlin1983,MooreRead1991}. Dashed guides indicate the $r^2$ and $r^6$ scalings.}
\label{fig:gr_loglog}
\end{figure}
\clearpage

\begin{figure}[!htbp]
\centering
\includegraphics[width=0.5\textwidth,height=0.5\textheight,keepaspectratio]{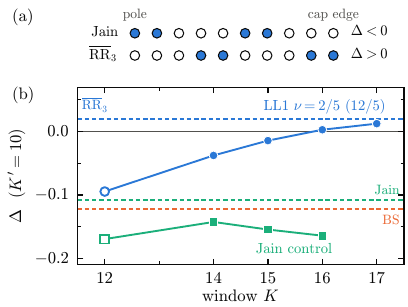}
\caption{\textbf{Conditional orbital polarization at LL1 $\nu=2/5$.}
For an inner cap of $K'$ orbitals, we define
\(
d(n)=\langle L_A\rangle_{n_A=n}-n(K'-1)/2
\)
and $\Delta=d(\nu K')$, measuring the departure from a uniform orbital distribution at fixed cap occupation.
(a) Jain and $\overline{\mathrm{RR}}_3$ root configurations illustrate opposite polarization directions.
(b) On the ten-orbital cap, the LL1 bootstrap polarization evolves from $-0.095$ at $K=12$ to $0.013$ at $K=17$, while the LLL Jain control remains near $-0.15$.
Exact finite-sphere references give $\widehat\Delta=-0.108$ for Jain and $-0.122$ for BS ($N_e=10$), compared with $+0.020$ for $\overline{\mathrm{RR}}_3$ ($N_e=12$), placing the bootstrap result closest to $\overline{\mathrm{RR}}_3$.
The sphere estimates account for the finite-size filling offset but retain curvature corrections; Laughlin benchmarks indicate a finite-size shift of $0.04$--$0.06$ on this cap (Methods).}
\label{fig:twofifths}
\end{figure}
\clearpage

\end{document}